# The effect of polymorphism on the structural dynamic and dielectric properties of plastic crystal water: A molecular dynamics simulation perspective.

Ioannis Skarmoutsos[1,*], Stefano Mossa[2] and Elvira Guardia[1,*]

[1] *Departament de Física, Universitat Politècnica de Catalunya, Campus Nord-Edifici B4-B5, Jordi Girona 1-3, Barcelona E-08034, Spain.*

[2] *Univ. Grenoble Alpes, CEA, CNRS, INAC-SyMMES, 38000 Grenoble, France.*

## Abstract

We have employed Molecular Dynamics simulations based on the TIP4P/2005 water model to investigate the local structural, dynamical and dielectric properties of the two recently reported body-centered-cubic and face-centered-cubic plastic crystal phases of water. Our results reveal significant differences in the local orientational structure and rotational dynamics of water molecules for the two polymorphs. The probability distributions of trigonal and tetrahedral order parameters exhibit a multi-modal structure, implying the existence of significant local orientational heterogeneities, particularly in the face-centered-cubic phase. The calculated hydrogen bond statistics and dynamics provide further indications of the existence of a strongly heterogeneous and rapidly interconverting local orientational structural network in both polymorphs. We have observed a hindered molecular rotation, much more pronounced in the body-centered-cubic phase, which is reflected by the decay of the fourth-order Legendre reorientational correlation functions and angular Van Hove functions. Molecular rotation, however, is additionally hindered in the high-pressure liquid compared to the plastic crystal phase. The results obtained also reveal significant differences in the dielectric properties of the polymorphs, due to the different dipolar orientational correlation characterizing each phase.

*Corresponding Authors: iskarmoutsos@hotmail.com, elvira.guardia@upc.edu

## I. Introduction

The exceptional and anomalous properties of water in a wide range of thermodynamic conditions keep stimulating the interest of the scientific community, due to its key role as a fundamental compound in numerous important chemical, biological and geological processes[1-5]. Despite long-lasting efforts new intriguing properties are still being described, and even the phase diagram of water, although systematically explored in the past[6-23], is far from being complete. A wide variety of polymorphs[24-26], including metastable states, has been reported for the solid phases of water. Interestingly, the formation of a plastic (or, as often called, rotator) crystal phase was quite recently demonstrated by Molecular Dynamics (MD) simulations[27-30]. In such a phase, although individual water molecules are able to rotate similar to the case of the liquid state, their centres of mass are precisely localized at the lattice sites of the underlying ordered structure[27,31-33].

The appearance of the plastic crystal phase of water at extreme pressure conditions has been first suggested theoretically, and an experimental verification has appeared to be challenging. Dolan et al.[34] have demonstrated experimentally using quasi-isentropic compression that at 400 K and pressures above 7 GPa liquid water can indeed undergo a first order phase transition to a solid, in just a few nanoseconds. The employed temporal and spatial resolution and the peaks intensity ratios, however, did not allow in that case to extract precise information about the resultant high-pressure phase and the transformation mechanism. As a consequence, it was not possible to distinguish between crystalline ice VII and the plastic crystal phase, as also revealed in the recent study of Gleason et al.[35]. In another experimental study, Schwager and Boehler[36] suggested the existence of a new ice phase, in the range from 20 to 42 GPa. Although the structure of this new phase was not identified, the melting curve was reported.

Computational molecular modelling can be an extremely useful tool in predicting the behaviour of water at extreme conditions, and can further guide and assist the design of experiments. By employing popular water potential models in classical MD and Monte Carlo (MC) simulations, in combination with free-energy calculation techniques, previous studies initially reported[27,28] the formation of a body-centered-cubic (bcc) plastic crystal phase and next of a face-centered-cubic (fcc) polymorph[29]. Aragones and Vega[29] showed that the bcc plastic crystal can transform into the fcc polymorph via a Martensitic phase transition[37] when heated at constant pressure, and that the transition temperature decreases on decreasing pressure. By using thermodynamic integration techniques, they also located the coexistence line for the bcc-fcc polymorphs, as well as for the ice VII - bcc plastic crystal phase, and for the fluid water - bcc plastic crystal. The corresponding fluid water-bcc-fcc, fluid water-ice VII-bcc and ice VII-bcc-fcc triple points located in this high-pressure region of the phase diagram were also estimated, for different water models, including TIP4P/2005, TIP4P, and SPC/E.

The plastic crystal phase of water can be distinguished from the ice and fluid phases on the basis of both thermodynamic and dynamic features. H. Tanaka and collaborators calculated reorientational and intermittent hydrogen bonds dynamics[31], together with the mean-squared displacements of water molecules in the bcc plastic crystal phase[27], by employing MD simulations of the TIP5P model. Those simulations revealed that the mean-squared displacement of water molecules slightly increases at very short times, eventually reaching a constant plateau value and immediately suggesting that translational motions are limited to vibrations around the lattice sites. The intermittent hydrogen bonding correlation function of the bcc plastic crystal phase also presented interesting features, exhibiting an initial decay similar to that of liquid water and finally converging to a finite non-zero value, indicating the recurrence of the hydrogen bonds[31]. In addition, the first order Legendre reorientational function decayed to zero, indicating that the water molecules were allowed to rotate in a genuine plastic phase.

All these findings clearly indicate the existence of the plastic crystal phase in simulated water, a robust conclusion further supported by coherent results produced by a wide variety of detailed force-fields representations[27-30], as well as toy models[27]. A unified and complete insight into the local structural order and related properties, together with an assessment of the effect of polymorphism upon these properties, however, has still not been provided. In this work, we therefore aim at shedding new light upon the differences in the local structural features and the related dynamical mechanisms in the two recently observed polymorphs of the plastic crystal phase of water, by employing MD simulations. We have paid particular attention to the local orientational order and the mechanisms governing the rotational dynamics in each polymorph. It has also been reported in the literature that the reorientational degrees of freedom exhibit considerable disorder in plastic crystals, and that molecular cooperativity and local orientational structural heterogeneities can reflect on dielectric properties[38,39]. For this reason, we have also further investigated the effect of the local structural cooperativity on static and dynamic dielectric properties of the polymorphs. The paper is organized as follows: computational details are presented in Section II, while our data and a comprehensive discussion of the results are presented in Section III. The main conclusions and final remarks are summarized in Section IV.

## II. Computational Details

We have performed classical Molecular Dynamics simulations to investigate the local structural, dynamic and dielectric properties of the bcc and fcc polymorphs of the plastic crystal phase of water, at pressure P=8 GPa and temperature T=440 K. We have used the TIP4P/2005[40] rigid potential model, which has been demonstrated to provide a realistic description of both normal properties and anomalies of water in a wide range of thermodynamic conditions. Note that the choice of a rigid non-polarizable force field in the present context is not completely obvious. Indeed, previous experimental works have reported that, for low temperatures and high pressures not exceeding 20 GPa, the molecular framework of water is preserved, and the pressure dependence of the O-H *intra*-molecular bond length is negligible[41,42]. At the substantially higher temperature of interest here, however, it is still not demonstrated experimentally to what extent intramolecular modes do couple to phase transformations at high enough pressures or if they can rather be safely neglected. To address this uncertainty, one could also employ force-field-based simulations including both intra-molecular degrees of freedom and polarizability effects, or even plain ab-initio simulations. Both these options go well beyond the scope of the present work, and we have rather made the choice of the TIP4P/2005 model[40], which assures the existence of plastic crystal phases at high temperature[40]. Breakthrough additional experimental work is needed to be final on this point.

The equations of motion for the rigid water molecules were expressed in terms of quaternions and integrated using a leapfrog-type Verlet algorithm[43], with an integration time step of 1 fs. A Nose-Hoover thermostat[44] and barostat[45] with relaxation times of 1.0 and 2.0 ps, respectively, were used to constrain temperature and pressure during the equilibration NPT-MD simulations. Tighter conditions, corresponding to relaxation times of 0.5 and 1.0 ps, respectively, were used following the equilibration procedure. A cut-off radius $r_c$=9.0 Å has been applied to the Lennard Jones interactions, and long-range corrections have been considered, while the Ewald summation technique has been used to account for the long-range electrostatic interactions. All simulations were performed using the DL_POLY package[46].

The target bcc plastic crystal phase was prepared as follows. First, an initial randomly disordered configuration consisting of $N_w$=432 water molecules comprised in a cubic box was prepared by using the *Packmol* tool[47]. The system was then equilibrated at ambient conditions by employing isothermal-isobaric (NPT) MD simulation techniques. Temperature and pressure of the system were next gradually increased until the desired conditions, T=440 K and P=8 GPa. At this stage, the configuration of the system corresponded to a metastable liquid phase which was subsequently propagated in a 10 ns (NPT) run. After about 4 ns, a liquid-plastic bcc phase transition was observed, which was indicated by a spontaneous jump of the density of the system from 1.626 to 1.661 g/cm$^3$. Very similar procedures to produce water bcc plastic crystal phases have already been reported in the literature[28].

In the fcc plastic crystal case, an initial fcc configuration containing 500 water molecules with random orientations was prepared in a cubic box. The system was then equilibrated for 10 ns in the (NPT)-ensemble with P=8 GPa and T=440 K. Aragones and Vega[29] have reported that the fcc plastic crystal phase stabilized by this protocol perfectly reproduces that obtained following the Martensitic phase transition of the bcc plastic crystal phase. (This takes place by increasing the c-edge of the bcc structure while performing anisotropic NPT-MD simulations.) Our method can therefore be considered as a faster and reliable approach to reproduce the fcc plastic crystal phase of water.

Following the equilibration simulations of the two polymorphs, additional 5 ns NPT-MD simulations were performed to calculate the density of each phase. The calculated density for the fcc plastic crystal phase was 1.678 g/cm$^3$. Next, NVT-MD simulations were performed at the calculated densities of the two phases, by initially slightly affinely deforming the cubic boxes, to match the appropriate density of each polymorph. The systems were again equilibrated for 2 ns and additional production runs of 2 ns were then employed to determine structural, dynamic and dielectric properties by direct analysis of the simulated trajectories. Note that both the free energy calculations performed in Ref.[29], as well as the very recent lattice-switch Monte Carlo simulations by Brukhno et al.[48], have suggested that at T=440 K and P=8 GPa the fcc plastic phase is more stable than that corresponding to the bcc arrangement. This result is consistent with the findings of Aragones and Vega, where it was noticed that in anisotropic NPT simulations the bcc plastic crystal phase transforms to the fcc structure.

## III. Results and Discussion

**Local Intermolecular Structure**

**Radial distribution functions.-** The local intermolecular structure in the two polymorphs can be characterized in further details in terms of the atom-atom pair radial distribution functions (rdf). The O-O, O-H and H-H rdfs corresponding to the bcc and fcc plastic crystal phases at 440 K and 8 GPa are compared in Figure 1. Differences in the local intermolecular structure can be clearly observed in these data. In the case of the O-O rdf, the first peak is located at 2.85 Å in both phases, exhibiting a larger intensity in the case of the fcc polymorph. On the other hand, the position of the first minimum is different. In the case of the fcc phase, the minimum is located at 3.55 Å, whereas the corresponding distance for the bcc phase is 3.95 Å. The coordination numbers corresponding to the first minimum of each O-O rdf are 12.1 and 14.0 for the fcc and bcc phase, respectively. These very large coordination numbers indicate a very dense packing in both polymorphs of the plastic crystal phase of water. The shape of the O-O rdfs at larger distances also reflects the long-range translational structural order, which is a typical feature of plastic crystal phases.

The shape of the O-H rdfs is also quite different in the two polymorphs. More in details, in the bcc phase we observe a very clear low-intensity peak located at 1.85 Å, whereas this feature is less pronounced and even similar to a shoulder in the fcc phase. An additional high intensity peak, located at 2.75 Å in the case of the fcc polymorph and at 2.85 Å in the bcc phase, follows, with an intensity which is much higher in the latter case. A third low-intensity peak is also observed at short distances in the case of the bcc polymorph, a feature that, in contrast, is not present in the fcc phase. The long-range translational order also exhibits substantial differences in the two cases, as in the shape of the oscillations at the larger distances, and still shows typical features of plastic crystal phases. Less pronounced differences are observed in the case of the H-H rdfs, with a first peak observed at 2.35 Å in both polymorphs, followed by a first minimum located at 3.05 Å for the bcc phase and at 3.15 Å for the fcc case, exhibiting long-range oscillations as in the case of the O-O and O-H rdfs.

As mentioned above, the most typical features of the considered plastic crystal polymorphs of water is the very dense packing, which is also reflected on the very large value of the coordination numbers at short intermolecular distances. To obtain more detailed insight about the packing of water molecules, the total O-O rdf was expressed in terms of the sum of the different contributions arising from nearest neighbors around a tagged molecule. The O-O rdf can indeed be expressed as a sum of the rdfs $g^n(r)$ [49,50], where the index n corresponds to the n$^{th}$ nearest neighbor around the reference molecule,

$$g^{OO}(r) = \sum_{n=1}^{\infty} g^n(r). \qquad (1)$$

The contributions to the overall O-O rdf corresponding to the first six nearest neighbors in the fcc and bcc plastic crystal phases are shown in Figure 2. Here, we can also observe that the sum $\sum_{n=1}^{6} g^n(r)$ adequately describes the shape of the O-O rdf up to the first peak located at 2.85 Å. In addition, the first six neighbors are packed in a sphere with a maximum radius of about 3.15 Å, where the $g^6(r)$ vanishes in both cases.

**Hydrogen bonding.-** The above very dense packing of water molecules at very short intermolecular distances in the plastic crystal phases obviously impacts the local hydrogen bonding structure, that we have investigated by using a simple commonly used geometric criterion[51,52]. According to the latter, a hydrogen bond between two water molecules exists if the interatomic distances are such that RO…O $\leq$ 3.6 Å, RH…O $\leq$ 2.4 Å, and the donor-acceptor angle H-O…O $\leq$ 30º. (Here the symbol ... corresponds to intermolecular distances). Note that numerous operative definitions for the hydrogen bond formation are present in the literature, together with studies devoted to their reliability[53]. Considering that geometric criteria indeed provide reasonable description of the hydrogen bonding structure and dynamics in a wide range of thermodynamic conditions[51,52,54,55], we decided to employ the above specific criterion consistently with our previous studies on water in several thermodynamic conditions[56,57]. We have estimated the relative fractions of water molecules forming 0 to 6 hydrogen bonds that we present in Table 1. In order to compare with the HB properties of liquid ambient water, the calculated fractions at 298.15 K and 1 bar using the TIP4P-2005 force field are also presented in Table 1. Interestingly, the fractions of molecules forming 3 and 4 hydrogen bonds are almost equal in both phases, whereas there is a quite significant (~11.5 and 12.4 % in the bcc and fcc phases, respectively) fraction of molecules forming 5 hydrogen bonds and even a very small fraction of molecules forming 6 hydrogen bonds. It is interesting to mention that in the ambient liquid phase the dominating fraction is the one corresponding to molecules forming 4 hydrogen bonds (Table 1), whereas the fraction of molecules forming 2 and 5 hydrogen bonds are also noticeably smaller in comparison with the values obtained for the two polymorphs of the plastic crystal phase of water. These findings clearly correlate with the very dense packing of water molecules in the investigated plastic crystal phases, also implying a possible distortion of the tetrahedral arrangement of the closest neighbours around any molecule.

We also studied the continuous dynamics of the hydrogen bonds formed in the two plastic crystal phases. The average hydrogen bond (HB) dynamics for pairs (i, j) of hydrogen-bonded molecules can be described in terms of the time correlation function (tcf)[58] :

$$C_{HB}(t) = \frac{\langle h_{ij}(0) \cdot h_{ij}(t) \rangle_{t^*}}{\langle h_{ij}(0)^2 \rangle} \qquad (2)$$

and the associated HB lifetime,

$$\tau_{HB} = \int_0^\infty C_{HB}(t) \cdot dt \qquad (3)$$

The variable $h_{ij}$ is defined as $h_{ij}(0) = 1$ if molecule j is hydrogen-bonded to molecule i at times 0 and t, and the bond has not been broken in the meantime for a period longer than $t^*$; $h_{ij}(0) = 0$ otherwise. The case $t^* = 0$ represents the so-called continuous definition, where the calculated tcf is the continuous one $C_{HB}^C(t)$ and the corresponding lifetime is the continuous lifetime $\tau_{HB}^C$. The $C_{HB}^C(t)$ for both plastic phases are presented in Figure 3. We can observe that the continuous HB dynamics is slower in the bcc phase compared to the fcc crystal. The breaking of hydrogen bonds, however, occurs on a much shorter time scale in comparison with ambient water. The calculated continuous HB lifetime $\tau_{HB}^C$ using the TIP4P-2005 force field at 298.15 K and 1 bar and the same HB criteria is about 0.6 ps, with continuous lifetimes $\tau_{HB}^C$ of 0.1 and 0.06 ps for the bcc and fcc phases, respectively. Such a finding indicates that the breaking of hydrogen bonds in the plastic crystal phase of water takes place much faster in comparison with the ambient liquid, which is reflected on the one order of magnitude difference between the obtained continuous HB lifetimes of the two different phases.

**Structural order parameters.-** Important implications of the observed very dense packing in the plastic crystal phases can be clearly identified on the translational and orientational local structural order parameters. Quite recently, Russo and Tanaka[59] introduced a structural descriptor, ζ, defined as the difference between the distance of the *closest* neighbor which is *not* hydrogen bonded to a reference molecule and that of the *farthest* hydrogen bonded molecule to the same reference molecule. In the liquid and supercooled phases, this structural descriptor can also be related to the separation between first and second solvation shell[59]. Here, where water is very densely packed in ordered structures, ζ provides a measure of the distortion of the local hydrogen bonding network around a tagged water molecule.

We present the calculated probability distributions of ζ for both plastic crystal phases in Figure 4. Both distributions are sensibly shifted toward negative values, with maxima at -0.10 and -0.15 Å for the bcc and fcc plastic crystal phase, respectively. These negative values indicate that the short-range local hydrogen bonding network is distorted due to the existence of non-hydrogen bonded closest neighbors at very short intermolecular distances. Note that this observation is consistent with the fact that, although as discussed above we can find up to 6 molecules at very short distances around a central one, a significant fraction of molecules form 2, 3 or 4 hydrogen bonds. As a consequence, in these very densely packed phases there is a significant fraction of non-hydrogen bonded neighbors which interpenetrate the short-range local structural network around each reference water molecule. Note that the structural descriptor $\zeta$ obtained for liquid ambient water using the same force field and HB definition is positive and has the value of 0.36 Å, indicating a clear separation between the first and second solvation shell in liquid water.

This particular local structural ordering in both polymorphs also strongly affects the local orientational ordering, that can be quantified in terms of the tetrahedral[60] ($q_4$) and trigonal[61] ($q_3$) orientational order parameters. The tetrahedral order parameter[60]

$$q_4 = 1 - \left\langle \frac{3}{8} \sum_{j=1}^{3} \sum_{k=j+1}^{4} \left( \cos\phi_{jik} + \frac{1}{3} \right)^2 \right\rangle \tag{4}$$

provides information about the extent to which a molecule and its four nearest neighbors adopt a tetrahedral arrangement. Here, $\phi_{jik}$ is the angle formed by the vectors $\vec{r}_{ij}$ and $\vec{r}_{ik}$ connecting the oxygen of the central molecule i to those of two of the four nearest neighbors j and k, and $q_4 = 1$ in a perfect tetrahedral network while $q_4 = 0$ in an ideal gas.

The $q_3$ order parameter [61,62] is in contrast defined in terms of the three closest neighbors of a given molecule i,

$$q_3 = 1 - \left\langle \frac{4}{7} \sum_{j=1}^{2} \sum_{k=j+1}^{3} \left( \cos\phi_{jik} + \frac{1}{2} \right)^2 \right\rangle \tag{5}$$

and, as in the previous case, it assumes a value of one for the perfectly order case, and of zero for a random distribution. The calculated probability distributions for $q_3$ and $q_4$ are presented in Figure 5.

The local orientational order in the two polymorphs is very different. More precisely, in the bcc plastic phase the distribution of both $q_3$ and $q_4$ is bimodal, exhibiting two peaks at $q_3 = 0.425$ and $q_3 = 0.955$ in the first case, and at $q_4 = 0.465$ and at $q_4 = 0.975$ in the other, indicating that structures with intermediate and strong orientational ordering coexist in the bcc plastic crystal phase. Also, the fraction of molecules exhibiting high trigonal order ($q_3 \geq 0.785$) is about 15%, whereas around 3% of the molecules only exhibit high tetrahedral order. This substantial absence of strong tetrahedral order, with a simultaneous non-negligible fraction of molecules exhibiting a strong trigonal order, is consistent with the above observation that the predominant fraction of hydrogen-bonded molecules exhibits 3 hydrogen bonds.

In contrast with the bcc structure, as it can be clearly seen from the calculated distributions, local orientational disorder is much more pronounced in the fcc plastic crystal. The distributions of $q_4$ and $q_3$ are in this case multimodal, and exhibit peaks at low and high $q_4$ and $q_3$ values, pointing to the existence of significant local orientational heterogeneities in this phase. Apart from the very different shape of the calculated distributions, the differences in the local orientational order in the two polymorphs are also reflected on the average $q_4$ and $q_3$ values obtained for each plastic crystal phase. The calculated average $q_4$ values for the fcc and bcc polymorphs are 0.338 and 0.437, respectively. Similarly, the average $q_3$ values for the fcc and bcc phases were 0.317 and 0.436, respectively. The lower average values of $q_4$ and $q_3$ obtained for the fcc phase, further verify our previous statement that local orientational disorder is much more pronounced in the case of the fcc polymorph.

The dynamics of the local orientational structural order in both polymorphs have also been investigated in terms of the corresponding average tcf [57]

$$C_{q_n}(t) = \frac{\langle \delta q_n(0) \cdot \delta q_n(t) \rangle}{\langle \delta q_n(0)^2 \rangle}, \quad \delta q_n(t) = q_n(t) - \langle q_n \rangle \quad n = 3, 4 \tag{6}$$

and the associated correlation times $\tau_{q_n}$,

$$\tau_{q_n} = \int_0^\infty C_{q_n}(t) \cdot dt \tag{7}$$

The tcfs for the fcc and bcc plastic crystal phases are presented in Figure 6. The local orientational order dynamics are slower in the bcc phase, consistently with the behavior of the continuous hydrogen bond dynamics discussed above. The correlation times of the trigonal order parameter are 0.024 and 0.015 ps for the bcc and fcc plastic crystal phases, respectively, and 0.032 and 0.018 ps for $q_4$. These very small relaxation times indicate the existence of a rapidly interconverting local orientational structural network in both polymorphs. Also, note that the correlation time for the tetrahedral order parameter in liquid water at ambient conditions has been estimated to be 0.4 ps [63], implying that the local orientational order dynamic are an order of magnitude faster than in the plastic crystal phase.

**Static structure factors.-** As mentioned in the introduction, an experimental observation of the plastic crystal phases of water has been demonstrated to be particularly problematic, a situation where computational molecular modelling can, in principle, provide useful help. In this perspective, we present at this point the simulated static structure factors, as one would measure by diffraction experiments with Neutrons and X-Rays. A direct calculation of the static structure factors can also confirm the symmetry of the two generated crystal phases. More in details, the Neutrons-weighted structure factor can be expressed as [50,64]

$$S^{Neutron}(q) = \left\langle \frac{N}{\sum_\alpha N_\alpha b_\alpha^2} \sum_\alpha \sum_\beta b_\alpha b_\beta S_{\alpha\beta}(\vec{q}) \right\rangle \quad , \quad q = |\vec{q}| \qquad (8)$$

where $b_\alpha$ is the coherent neutron scattering length for species α. Here, the brackets indicate a spherical average over wave vectors of modulus q, and the partial static structure factors involving species α and β are defined as

$$S_{\alpha\beta}(\vec{q}) = \frac{(1+\delta_{\alpha\beta})}{2N} \rho_\alpha(\vec{q}) \rho_\beta^*(\vec{q}) \, , \qquad (9)$$

$$\rho_\alpha(\vec{q}) = \sum_{l=1}^{N_\alpha} \exp(i \cdot \vec{q} \cdot \vec{r}_l) \, , \qquad (10)$$

with $\vec{r}_l$ the instantaneous vector position of atom $l$.

In the case of X-Ray scattering the structure factor can similarly be expressed as

$$S^{X-Ray}(q) = \sum_{\alpha}\sum_{\beta} S^X_{\alpha\beta}(\vec{q}) \qquad (11)$$

where:

$$S^X_{\alpha\beta}(\vec{q}) = \frac{f_\alpha(q)f_\beta(q)}{\sum_\alpha \chi_\alpha f_\alpha^2(q)} S_{\alpha\beta}(\vec{q}) \qquad (12)$$

Here, $\chi_\alpha$ is the fraction of particles of species α, and $f_\alpha(q)$ and $f_\beta(q)$ are the appropriate q-dependent atomic form factors for species α and β, which can be approximated with a series of Gaussian functions of q

$$f_\alpha(q) = \sum_{i=1}^{4} a_i \exp\left[-b_i\left(\frac{q}{4\pi}\right)^2\right] + c \qquad (13)$$

The appropriate values for the parameters $a_i$, $b_i$, c can be found in the International Tables for Crystallography. Note that, especially in the low-q regime, the above direct method is a better choice than that relying on the Fourier transform of the radial distribution function, which suffers from both cut-off ripple artifact of g(r) and finite size effects [65].

We show the Neutrons and X-Ray static structure factors calculated for both the fcc and bcc plastic crystal phases in Figure 7, on a linear-log scale. This representation allows to underline a quite complex general shape of the S(q). This shows clear modulations, with peaks broadening recalling the case of amorphous-like media, probably due to the disordered rotational configuration of the *atoms*. This background structure, however, is modified by a superimposed series of well-defined Bragg peaks, corresponding to the ordering of the *molecular* centers of mass at the underlying crystal lattices sites. The positions of these peaks immediately provide us the symmetry of the corresponding crystal. Indeed, in the fcc polymorph's static structure factor the peak positions are $q_1$=2.62 and $q_2$=3.03 Å$^{-1}$. The corresponding peak positions in the bcc polymorph's static structure factor are $q_1$=2.69 and $q_2$=3.8 Å$^{-1}$. The ratios $q_2/q_1$=2/3$^{1/2}$=1.15 and $q_2/q_1$ =2$^{1/2}$=1.41 are therefore obtained in the cases of the fcc and bcc plastic crystals, respectively. These ratios exactly correspond to those expected for the smallest allowed reflections (111) and (200) in the fcc polymorph and (110), (200) in the bcc one. Also, note that these peaks are more intense in the low-q regime, with an intensity steadily decreasing as the value of q increases. This is exactly what one should expect when genuine long-range structural order is present and increases at length scales larger than that associated to the rotational disorder.

We conclude this Section with a last observation. As mentioned above, an experimental observation of the plastic crystal phases of water has been demonstrated to be particularly problematic[34-36] and computational molecular modelling is recognized to be an effective tool to provide useful guide to lift some of these difficulties. In this sense, our data can represent a useful benchmark for future studies based on diffraction experiments with Neutrons and X-Rays.

## B. Translational-Rotational Dynamics

We now turn our attention to the dynamics of the plastic crystal phases which for the translational degrees of freedom motions are limited to librations around the lattice sites. The translational dynamics can be characterized in terms of the tcfs of the atomic velocities

$$C_v^i(t) = \frac{\langle \vec{v}_i(0) \cdot \vec{v}_i(t) \rangle}{\langle \vec{v}_i(0)^2 \rangle} \tag{14}$$

with the associated spectral densities, $S_v^i(\omega)$, calculated by performing a Fourier transform

$$S_v^i(\omega) = \int_0^\infty \cos(\omega \cdot t) \cdot C_v^i(t) \cdot dt \tag{15}$$

Here, the $S_v^i(\omega)$ have been calculated by numerical integration using a Bode rule after convolution of the calculated velocity tcfs with a Hanning window. The calculated water oxygen and hydrogens velocity tcfs and the corresponding spectral densities are presented in Figures 8 and 9, respectively. Based on the data of Figure 8, we observe that the atomic translational dynamics are very fast, with a tcfs vanishing at about 0.2 ps after a series of negative local minima on the time scale range 0.025-0.15 ps, which reflect the librational motions around the lattice sites. Altogether, differences in the behaviour of the polymorphs are tiny in these data.

Relative variations in the two systems are more evident by analysing the shapes of the calculated spectral densities, as shown in Figure 9. Interestingly, the spectral densities associated to oxygen do not exhibit the low frequency peak around 60 cm$^{-1}$ observed in the liquid phase[62], while presenting a peak at 165 cm$^{-1}$ and 180 cm$^{-1}$ for the bcc and fcc systems, respectively. These features are complemented by a shoulder at around 355 cm$^{-1}$ in both phases. In addition, the spectral density associated to the hydrogens exhibit a wide peak centered around 485 cm$^{-1}$ in the bcc plastic crystal phase, while a small peak at 353 cm$^{-1}$ is followed by a shoulder at about 485 cm$^{-1}$ in the case of the fcc polymorph. Another, higher frequency, shoulder is also observed for both plastic crystal phases at around 830 cm$^{-1}$. Note that the blue-shifting of the librational frequencies in the plastic crystal phases compared to liquid water could be considered as an effect of the very dense packing of water molecules at very short distances, as already stressed in the discussion above.

As previously mentioned, while translations are over-constrained in plastic crystals, individual water molecules can in contrast rotate around the lattice sites. We can efficiently characterize rotational dynamics in terms of the well-known Legendre reorientational tcfs of adapted molecular vectors

$$C_{\ell R}(t) = P_{\ell} \left\langle \vec{u}_i(0) \cdot \vec{u}_i(t) \right\rangle \qquad (16)$$

Here, $\vec{u}_i$ is a normalized vector associated to molecule i, and $P_{\ell}$ is a Legendre polynomial of order $\ell$. While reorientational dynamics in condensed phases has been mostly associated to the behaviour of first and second-order Legendre polynomials, previous studies[66,67] also including our previous investigation of confined systems in narrow pores[68], have demonstrated that important features can also be reflected on the behaviour of higher-order Legendre reorientational tcfs. We have therefore calculated the latter up to the fourth order for the O-H vector of the water molecules, and the results are shown in Figure 10 for the two plastic crystal phases. These data reveal that the fourth-order Legendre tcf for the O-H vector reaches at long-time scales a non-zero plateau value, whereas Legendre tcfs of order 1, 2, and 3 all decay to zero. Interestingly, our previous studies of molecules confined in very small nanopores[68], as well as other investigations of model diatomic supercooled liquids[66] and liquid symmetric dumbbells[67], have demonstrated that such a behaviour is an indication of hindered rotation. Note that the height of the plateau is more pronounced in the bcc plastic crystal phase, implying that rotation dynamics in this polymorph are more hindered than in the fcc case.

Deeper insight on the rotational motions of water molecules in the two polymorphs comes from an analysis of the self angular Van Hove correlation functions for the O-H vector, calculated as[68,69]

$$G(\theta,t) = \frac{2}{N \sin\theta} \sum_{i=1}^{N} \left\langle \delta\left[\theta - \cos^{-1}(\vec{u}_i(0) \cdot \vec{u}_i(t))\right] \right\rangle \tag{17}$$

The calculated values for $1/2 \cdot \sin\theta \cdot G(\theta,t)$ at the indicated values of t and as a function of theta are shown in Figure 11 for both polymorphs. Based on these data, it is possible to confirm that indeed the rotation of the O-H vector is additionally hindered in the bcc plastic crystal phase compared to the fcc crystal. In particular, at long times in the bcc phase the data are characterized by the appearance and development of two distinct shoulders, located at about 20 and 160 degrees. This behaviour is in contrast much less pronounced in the case of the fcc plastic crystal phase. This finding clearly indicates that, occasionally, water molecules perform jump-like orientational flips of nearly 180°, leading to the observed very slow relaxation of the fourth-order reorientational correlators. Note that the proposed orientational flips must be rare, otherwise sharp peaks would be observed in the distribution functions.

The difference in the rotational dynamic behavior of water molecules in the two plastic crystal phases can be additionally highlighted by calculating the angular mean-squared displacements together with the corresponding rotational diffusion coefficient, through the Einstein relation[70]

$$D_R = \lim_{t \to \infty} \frac{1}{4t} \left\langle \Delta\varphi^2(t) \right\rangle \tag{18}$$

The time-dependence of the calculated average mean-squared angular displacements for water molecules in the fcc and bcc plastic crystal phases is presented in Figure 12. From this Figure it can be clearly seen that water molecules in the fcc plastic crystal phase rotate faster than in the bcc system, with rotational diffusion coefficients for the O-H vector $D_R$ = 1.15 rad$^2$/ps and $D_R$ = 0.92 rad$^2$/ps for the fcc and bcc phases, respectively. Note that these values are significantly higher than those reported experimentally and theoretically for liquid water at 400 K and pressures close to 3 GPa[71], indicating that rotation is more hindered in the high pressure liquid phase than in the plastic crystal polymorphs.

## C. Dielectric Properties

Also based on previous studies focusing on liquid and supercritical water[72,73], we have performed an analysis of the static and dynamic dielectric properties of the two plastic crystal polymorphs. The static dielectric constant of the system can be calculated as

$$\varepsilon_0 = \varepsilon_\infty + 3yG_K \qquad (19)$$

where $\varepsilon_\infty$ is the dielectric constant at optical frequencies, approaching the value 1 at high frequencies because we assumed rigid non-polarizable molecules, and $y$ is the dimensionless dipolar strength defined as

$$y = \frac{4\pi\rho\vec{\mu}^2}{9k_BT} \qquad (20)$$

with $\rho$ is the number density of water and $\vec{\mu}$ the molecular dipole vector. In addition, $G_K$ is the finite system Kirkwood factor given by

$$G_K = \frac{\langle \vec{M}^2 \rangle}{\langle N\vec{\mu}^2 \rangle} \qquad (21)$$

where $\vec{M} = \sum_i \vec{\mu}_i$ is the total dipole moment of the system. The infinite system Kirkwood correlation factor $g_K$ can be calculated from the equation:

$$g_K = \frac{2\varepsilon_0 + \varepsilon_\infty}{3\varepsilon_0} G_K \tag{22}$$

The frequency dependent dielectric constant can be expressed as

$$\varepsilon(\omega) = \varepsilon'(\omega) - i \cdot \varepsilon''(\omega) \tag{23}$$

where $\varepsilon'(\omega)$ and $\varepsilon''(\omega)$ are the real and imaginary parts that can be calculated according to

$$\varepsilon'(\omega) = \varepsilon_0 - (\varepsilon_0 - \varepsilon_\infty) \cdot \omega \cdot \int_0^\infty \Phi(t) \cdot \sin(\omega \cdot t) \cdot dt \tag{24}$$

and

$$\varepsilon''(\omega) = (\varepsilon_0 - \varepsilon_\infty) \cdot \omega \cdot \int_0^\infty \Phi(t) \cdot \cos(\omega \cdot t) \cdot dt \tag{25}$$

where $\Phi(t)$ is the total dipole tcf

$$\Phi(t) = \frac{\langle \vec{M}(0) \cdot \vec{M}(t) \rangle}{\langle \vec{M}^2 \rangle} \tag{26}$$

Two observations are in order. First, the Debye relaxation time can be determined from the relation

$$\tau_D = \lim_{\omega \to 0} \left[ \frac{\varepsilon_0 - \varepsilon(\omega)}{i \cdot \omega \cdot (\varepsilon(\omega) - \varepsilon_\infty)} \right] \tag{27}$$

Second, in the present case we have found that the time decay of $\Phi(t)$ can be represented by a single exponential decay function:

$$\Phi(t) = \exp(-t/\tau) \tag{28}$$

where $\tau$ is the exponential time decay constant, such that the frequency dependent dielectric constant can be expressed as:

$$\varepsilon(\omega) = \varepsilon_\infty + \frac{\varepsilon_0 - \varepsilon_\infty}{1 + i \cdot \omega \cdot \tau} \tag{29}$$

Static and dynamic dielectric properties of the two plastic crystal phases of water determined based on the above formalism are summarized in Table 2. These data indicate significant differences in the two polymorphs. The static dielectric constant of the fcc plastic crystal phase is 92.3, whereas it is 65.9 in the case of the bcc phase, a discrepancy indicating that the local dipolar arrangements are very different in each polymorph and determine very different collective total dipole moment arrangements. This conclusion is further supported by the observation that both finite and infinite system Kirkwood factors are also very different in the two phases, as it is evident from Table 2.

We further characterized the local dipolar arrangements in the two phases in terms of the average value of the cosine of the angle formed by the dipole moment vectors of two water molecules as a function of their distance r, $\langle \cos\theta_\mu(r) \rangle$, shown in Figure 13 a). These data clearly support the conclusion of the existence of significant differences in the dipolar arrangements of water molecules in the two polymorphs. The fluctuations in the radial dependence of the cosine, which are much more pronounced in the case of the fcc plastic crystal phase, are also reflected on the radial dependence of the Kirkwood factor, $G_K(r)$, as it can be seen in Figure 13 b). Our conclusion is therefore that these fluctuations play an important role in determining the collective total dipole moment and the static dielectric constant of the polymorphs. It is worth to mention that, for both plastic phases, the fluctuations in $G_K(r)$ extend to higher distances than those observed in liquid water[74].

Real and imaginary parts of the frequency dependent dielectric constant are shown in Figures 14 a) and b) for both polymorphs. The shape of the dielectric spectra shows that the mechanisms underlying the dielectric relaxation are not very different in the two phases, a conclusion further corroborated by the fact that the calculated Debye relaxation times in the two crystalline phases are not very different, amounting to 0.95 ps and 0.85 ps in the fcc and bcc, respectively. We also note that the corresponding exponential time decay constants, $\tau$, presented in Table 2 are almost identical to the calculated Debye relaxation time, confirming that the relaxation of the collective dipole moment of both polymorphs is controlled by a single exponential relaxation process. The resulting Cole-Cole plots[71] for the two plastic crystal phases are also presented in Figure 14 c). The loops observed in the high frequency region, and the deviation from a semi-circular behaviour in this region in both polymorphs clearly reflects the existence of librational motions discussed above in terms of the atomic velocity tcfs and the corresponding spectral densities.

## IV. Conclusions

We have performed Molecular Dynamics simulations of two different polymorphs of solid water, the fcc and bcc plastic crystal phases, at 440 K and 8 GPa based on the TIP4P/2005 classical potential model. The simulations have revealed a dense short-range packing in both phases and significant differences in the shape of the atom-atom radial distribution functions. The analysis of the degree of short-range translational order, and of the local hydrogen bonding network in both polymorphs has also allowed to identify a significant fraction of non-hydrogen-bonded neighbors around each individual water molecules, which penetrate its short-rang environment and create a strongly heterogeneous and rapidly interconverting local structural network in both polymorphs. We have also found evidences of the existence of significant local orientational heterogeneities, reflected on the shape of the probability distributions of the orientational order parameters of water, particularly in the fcc plastic crystal phase.

Neutrons and X-Ray weighted static structure factors were also determined, in the hope to provide useful guidelines for future experimental work. The results obtained have revealed that the shape of the structure factor recalls that typical of an amorphous-like medium, interpreted in terms of the random atomic orientations. This background is decorated by a series of superimposed Bragg peaks, determined by the long-range structural order associated to the molecular centers of mass.

The results obtained also point to the existence of hindered molecular rotations, much more pronounced in the body-centered-cubic phase, which are reflected by the decay of the fourth order Legendre reorientational correlation functions and of the angular Van Hove functions. We have found, however, that rotations are more hindered in the high-pressure liquid phase than in the plastic crystal. On the other hand, the very dense packing of water molecules at very short distances causes a blue-shifting of the librational frequencies in the plastic crystal phases compared to liquid water, as it is revealed by the shape of the calculated atomic velocity tcfs and by the corresponding spectral densities.

The present work has also revealed that the calculated static dielectric constant of the two polymorphs are significantly different. This difference can be rationalized by noticing that the local dipolar arrangements are very different in the investigated polymorph, determining very different collective total dipole moment and Kirkwood factors in the two cases. An analysis of the shape of the dielectric spectra actually reveals that the mechanisms controlling the dielectric relaxation are not very different in each phase, and that the relaxation of the collective dipole moment can be described in terms of a single exponential relaxation process in both systems. Finally, the loops observed in the high frequency region of the Cole-Cole plots, together with a significant deviation from a semi-circular regime in this region clearly reflects the existence of librational motions in both polymorphs, a conclusion also supported by an analysis of the translational dynamics.

## Acknowledgements

I. S. acknowledges financial support from the group of Professor Elvira Guardia at the Department of Physics of the Technical University of Catalonia (UPC), which covered the expenses of his visiting scientist position in the same institution during the periods June-July 2016 and April-May 2018. E.G. acknowledges financial support from the Ministerio de Economía y Competitividad of Spain, Grant FIS2015-66879-C2-1-P (MINECO/FEDER).

# TABLES

**Table 1:** Calculated fractions of water molecules forming 0-6 hydrogen bonds in the investigated plastic crystal phases of water at 440 K and 8 GPa. Comparison with the results obtained for liquid water at 298.15 K and 1 bar is also provided.

| Phase | $f_0$ (%) | $f_1$ (%) | $f_2$ (%) | $f_3$ (%) | $f_4$ (%) | $f_5$ (%) | $f_6$ (%) |
|---|---|---|---|---|---|---|---|
| Plastic bcc | 0.104 | 2.034 | 13.216 | 35.597 | 36.576 | 11.47 | 1.003 |
| Plastic fcc | 0.112 | 2.18 | 13.85 | 35.143 | 34.778 | 12.397 | 1.54 |
| Liquid Water | 0.038 | 0.948 | 8.288 | 31.873 | 53.738 | 5.087 | 0.028 |

**Table 2:** Calculated static and dynamic dielectric properties of the investigated plastic crystal phases of water at 440 K and 8 GPa.

| Phase | $\mu$ (D) | $G_K$ | $y$ | $\varepsilon_\infty$ | $\varepsilon_0$ | $g_K$ | $\tau_D$ (ps) | $\tau$ (ps) |
|---|---|---|---|---|---|---|---|---|
| Plastic bcc | 2.305 | 3.19 | 6.78 | 1 | 65.9 | 2.14 | 0.85 | 0.87 |
| Plastic fcc | 2.305 | 4.44 | 6.85 | 1 | 92.3 | 2.98 | 0.95 | 0.94 |

**FIGURES**

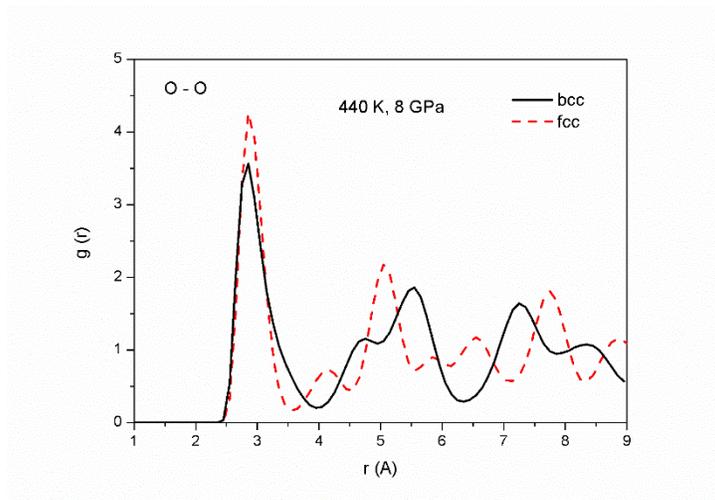

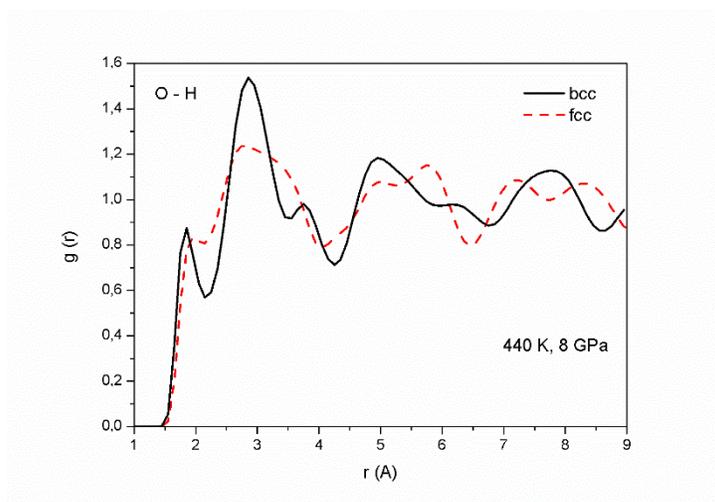

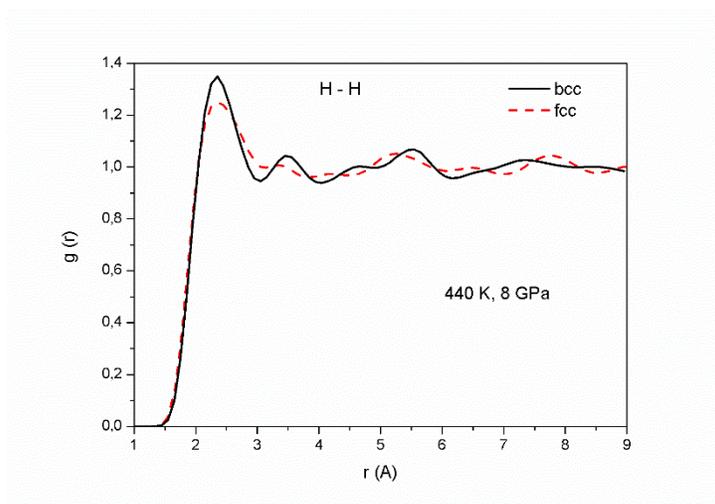

**Figure 1**

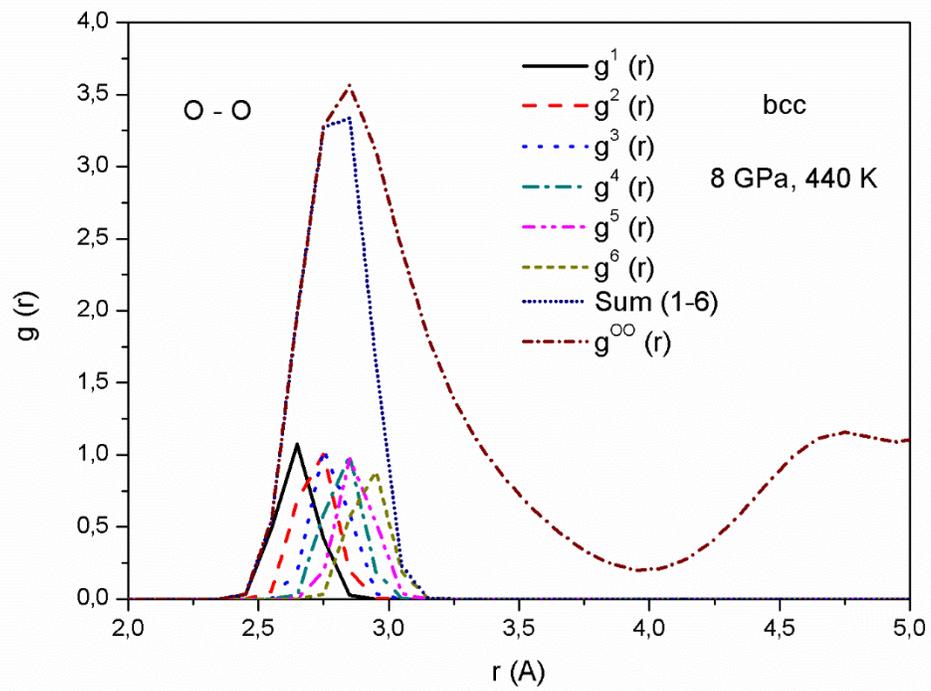
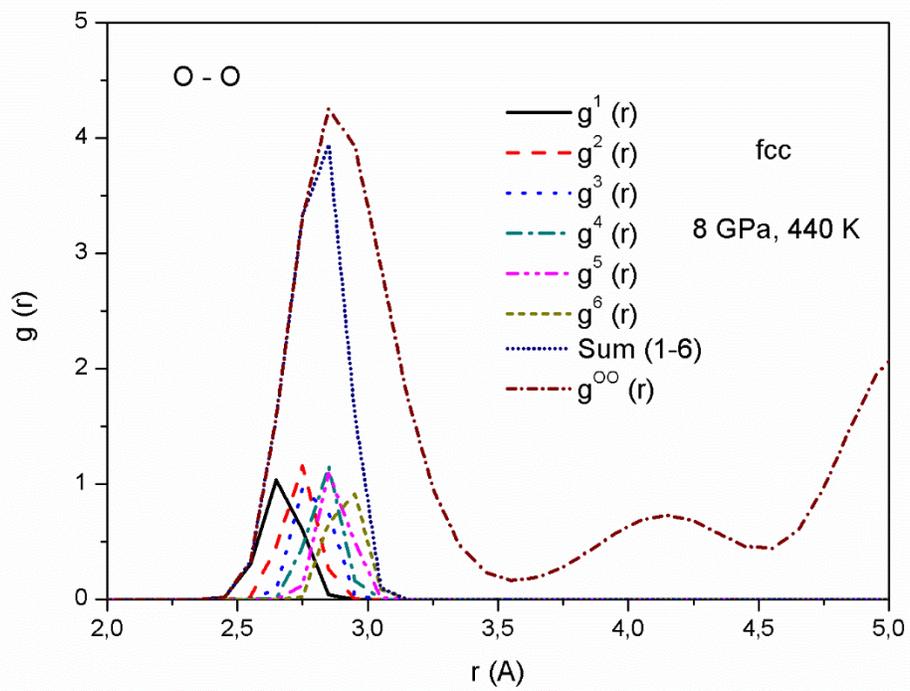

**Figure 2**

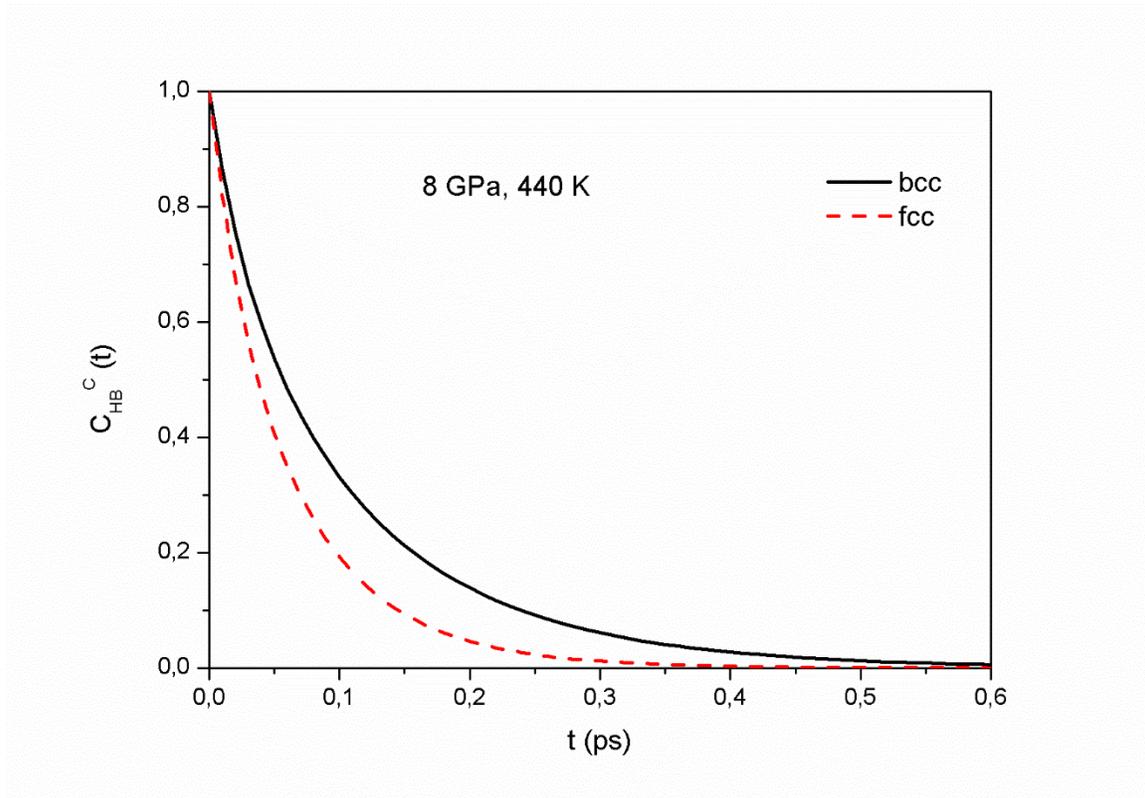

**Figure 3**

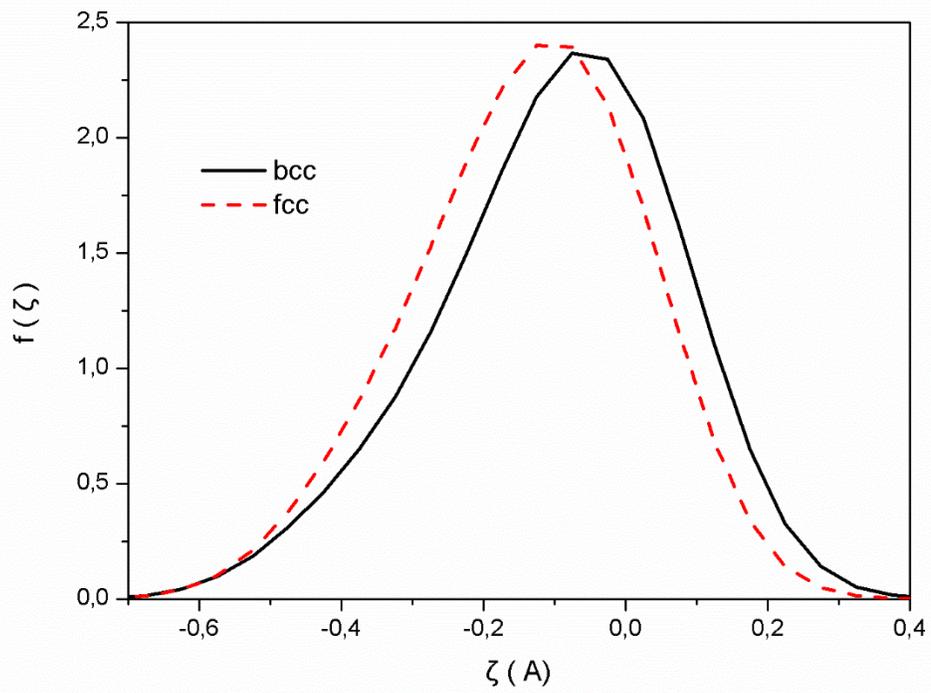

**Figure 4**

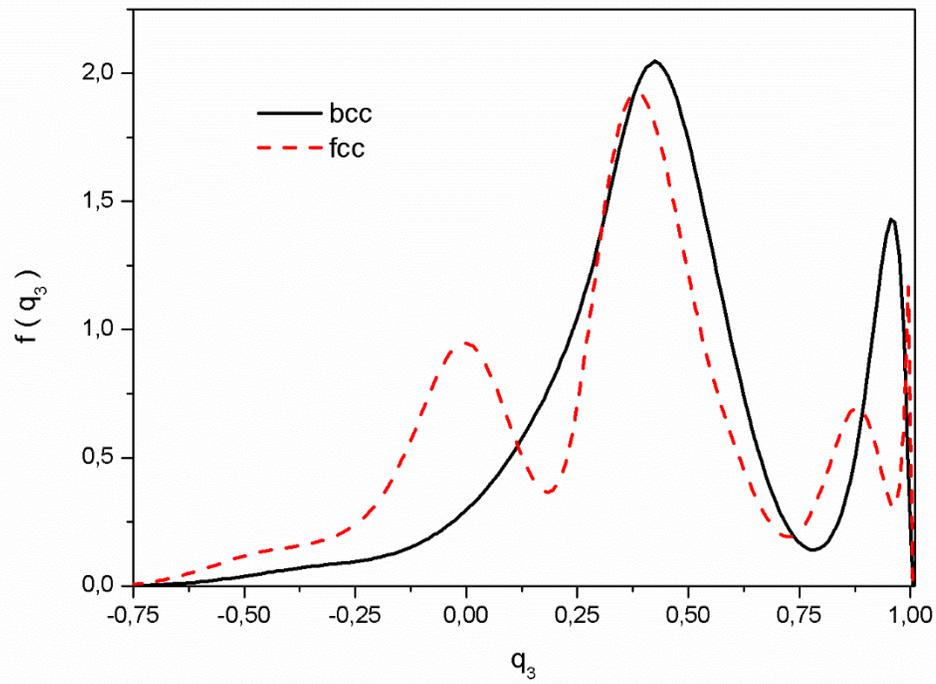

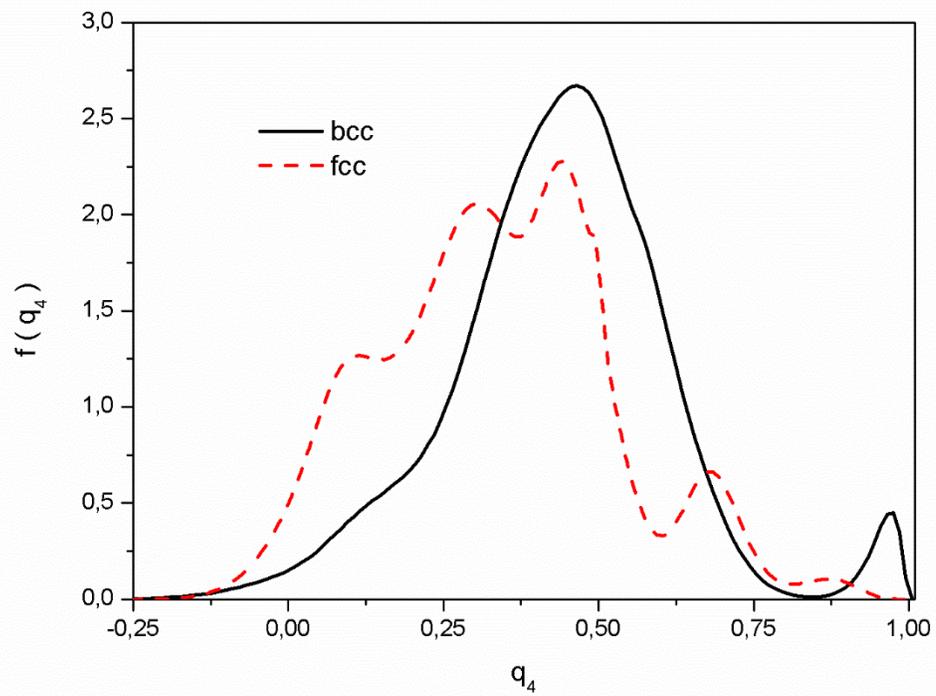

**Figure 5**

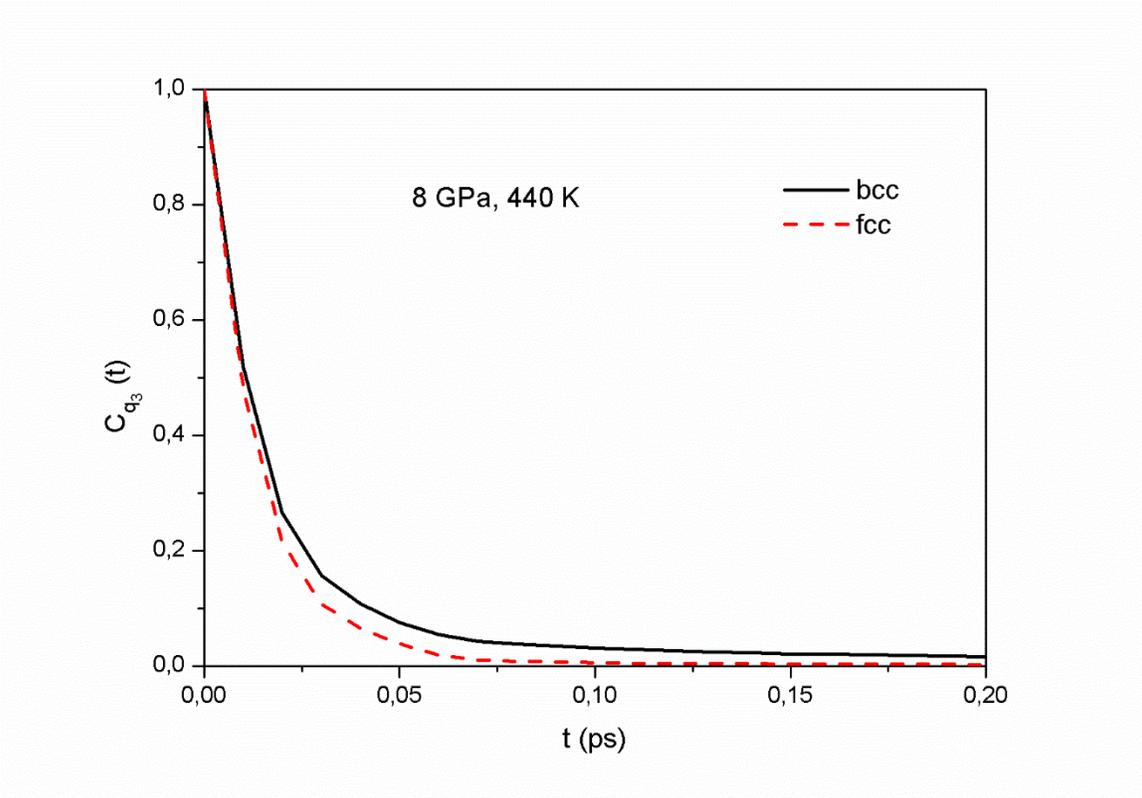

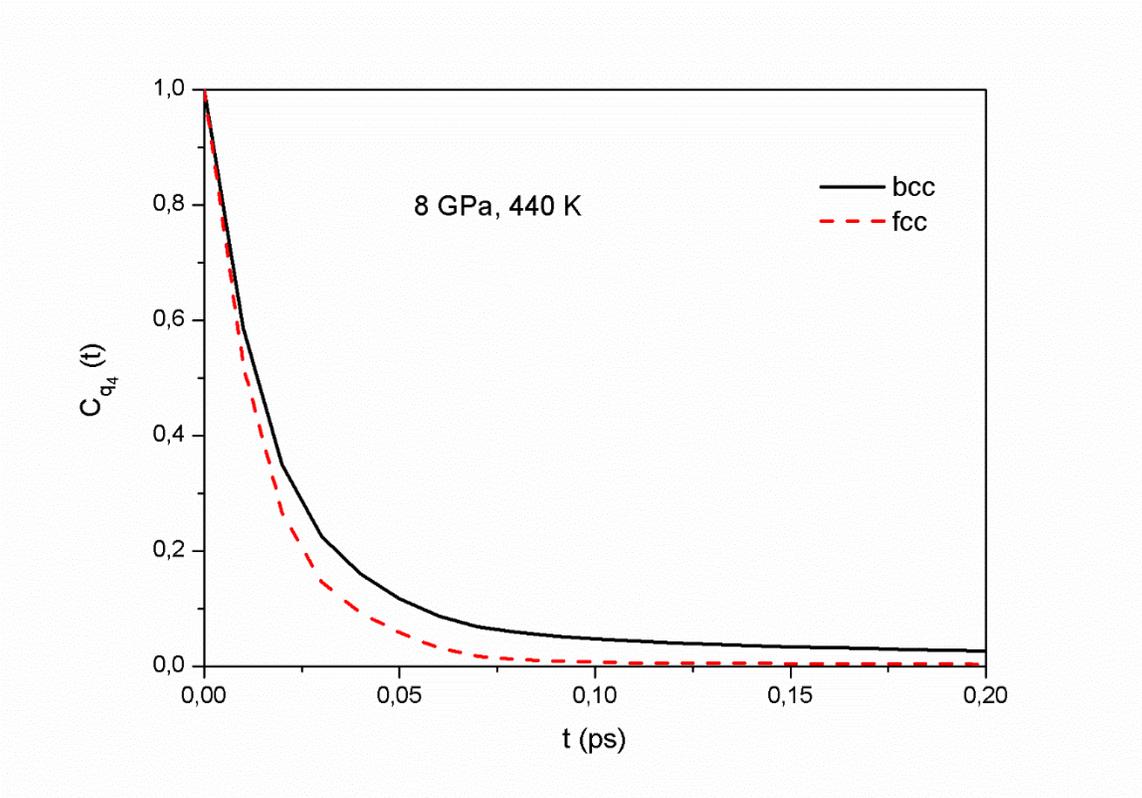

**Figure 6**

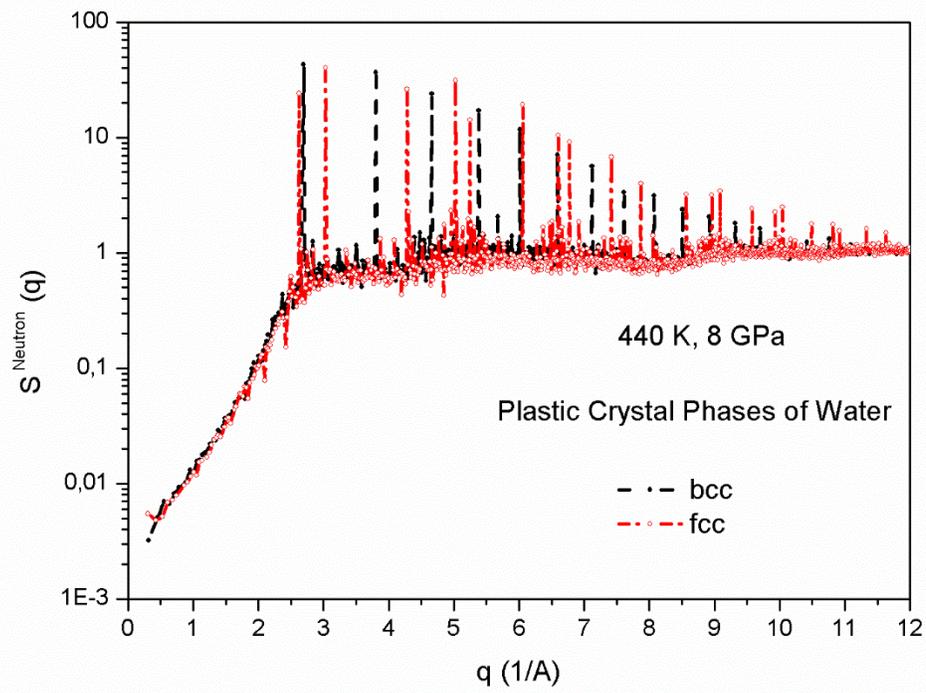

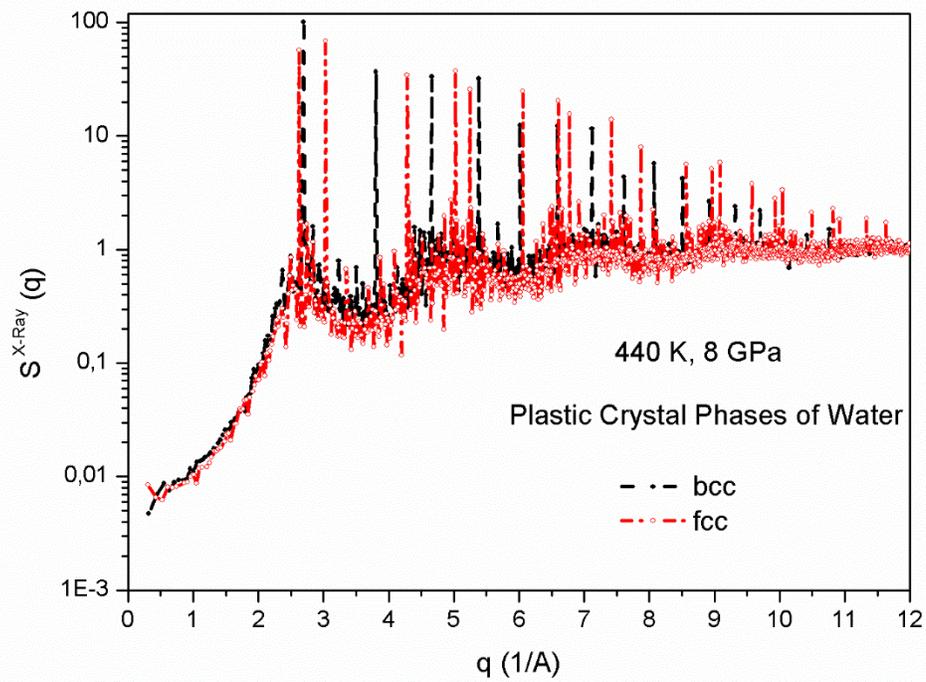

**Figure 7**

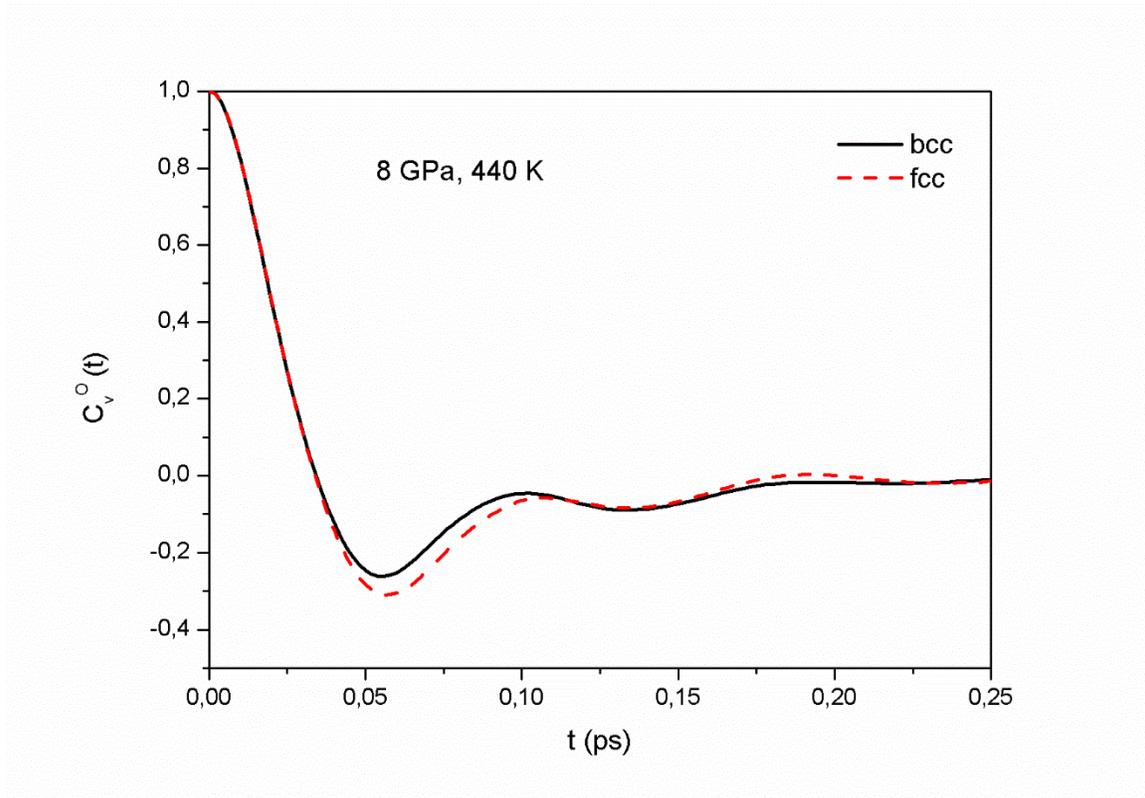
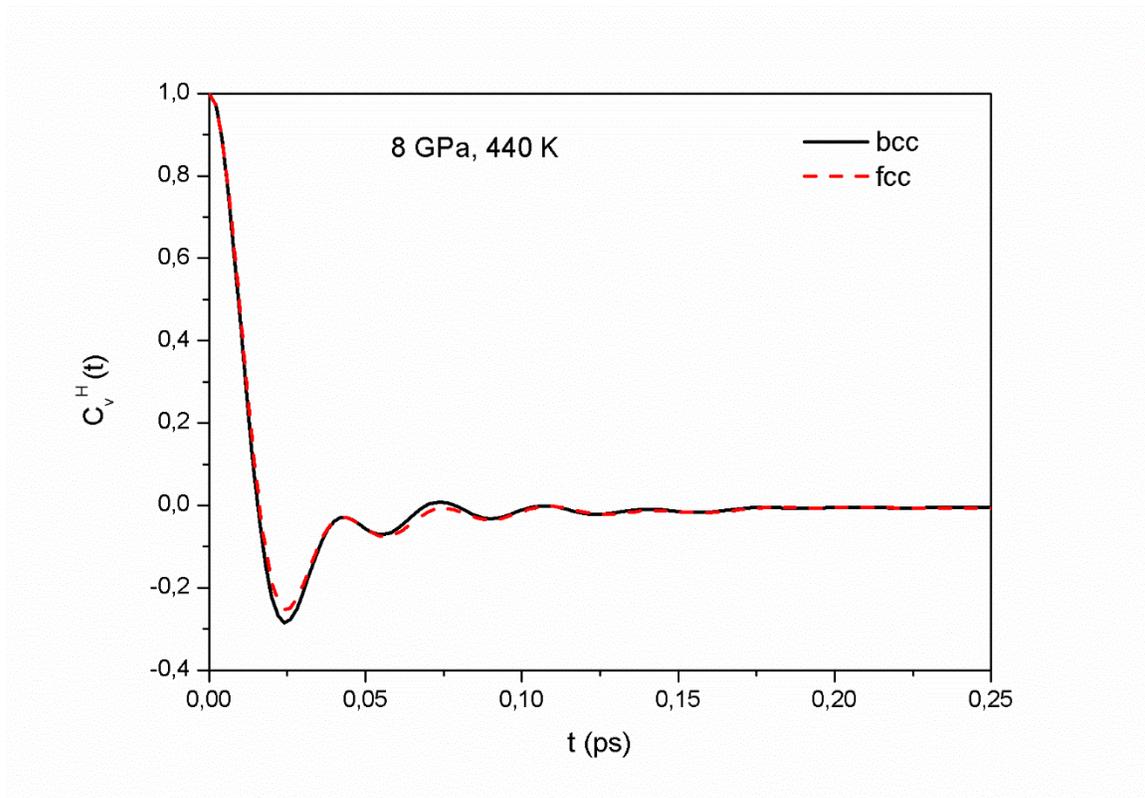

**Figure 8**

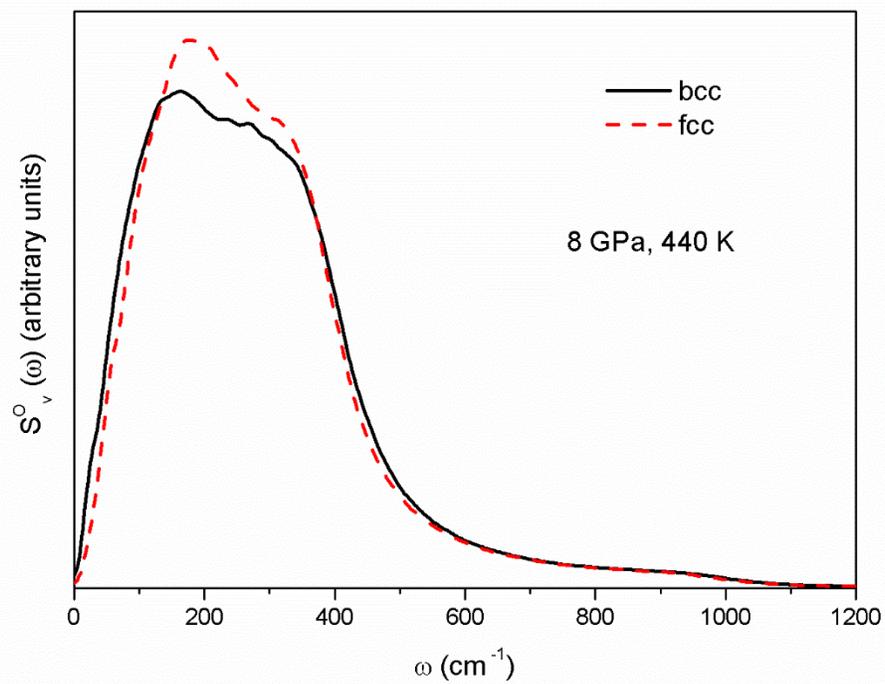

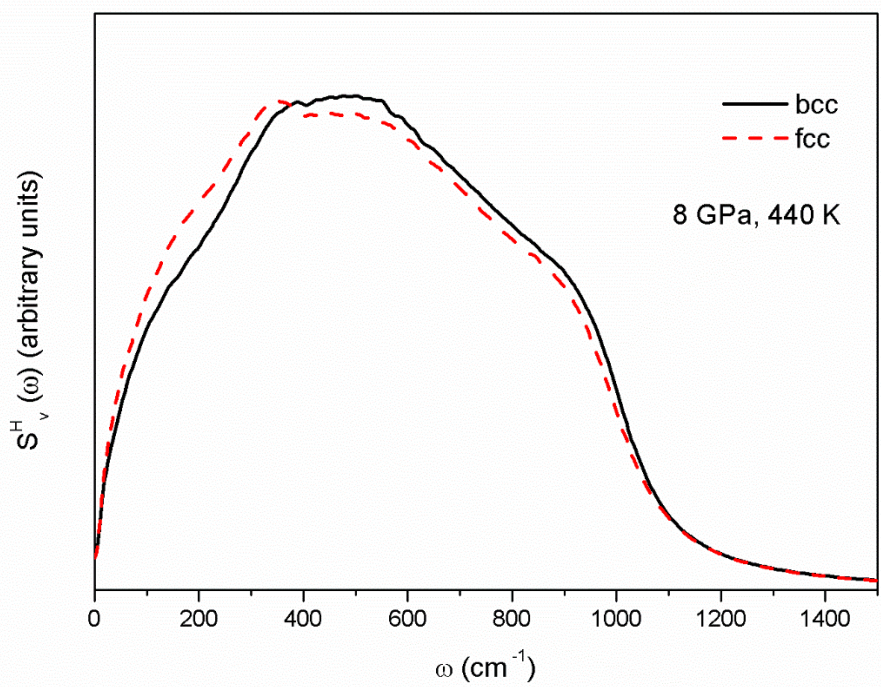

**Figure 9**

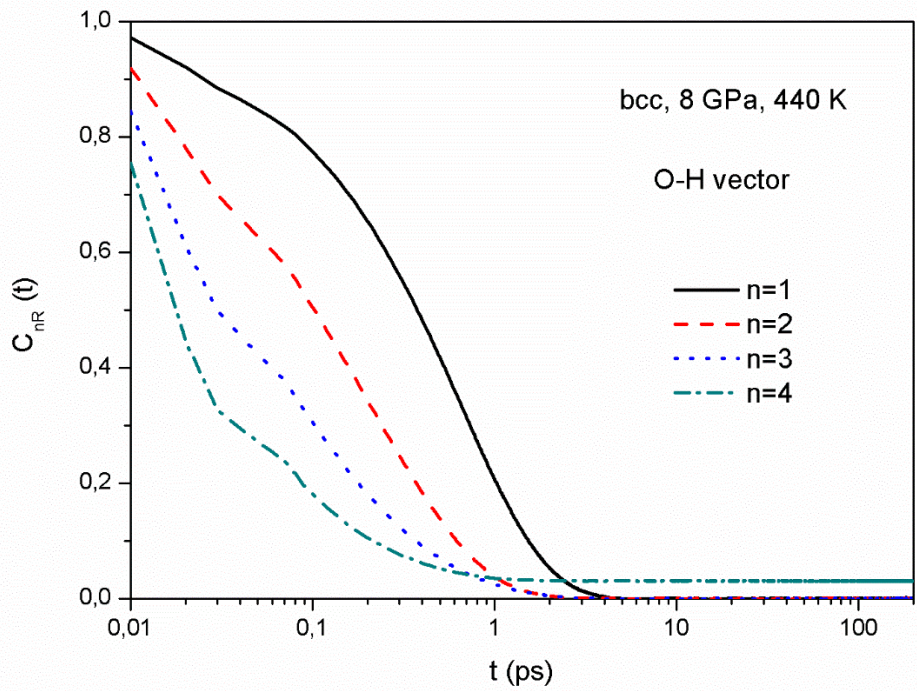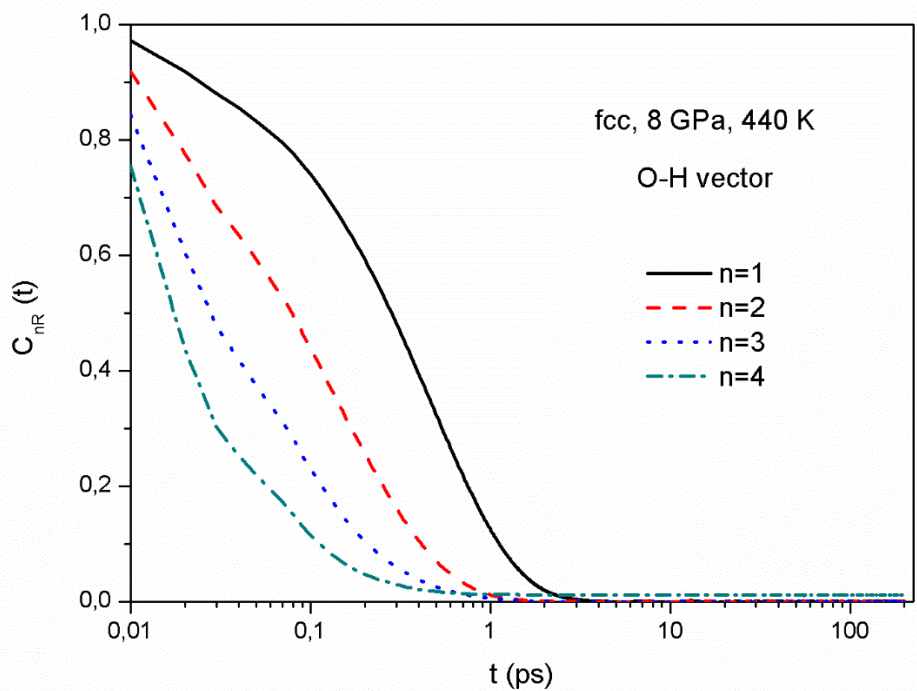

**Figure 10**

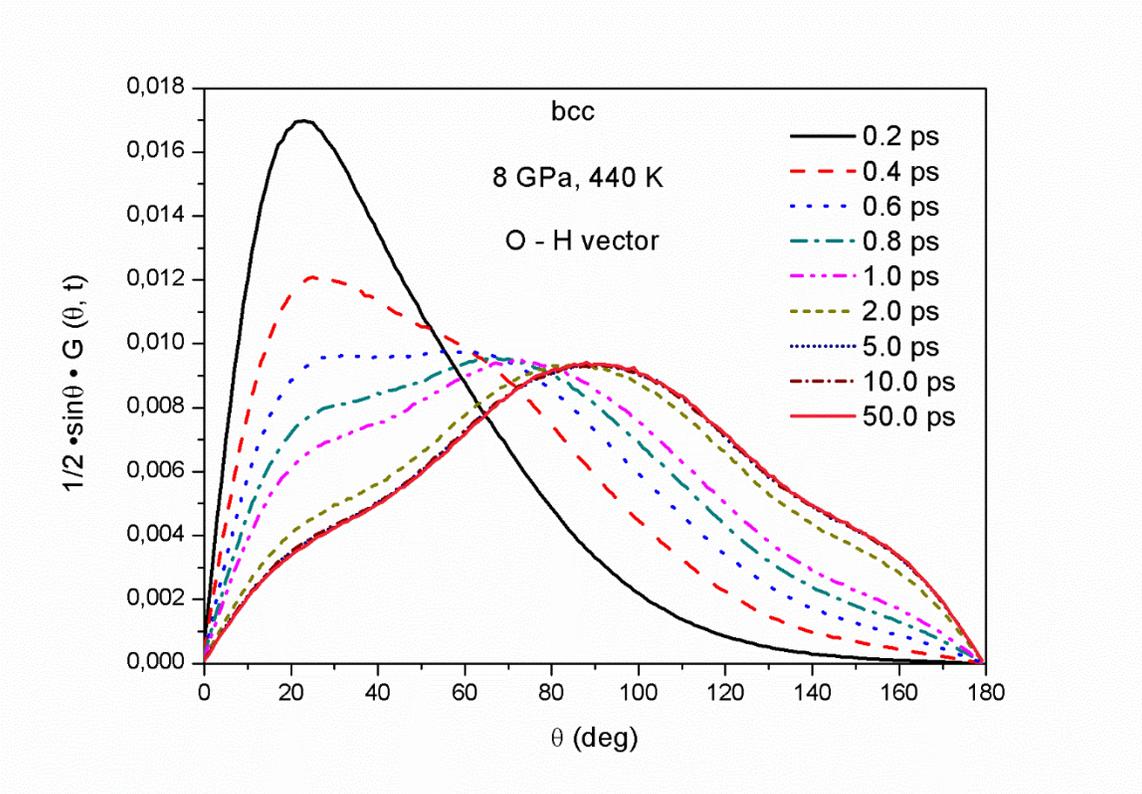

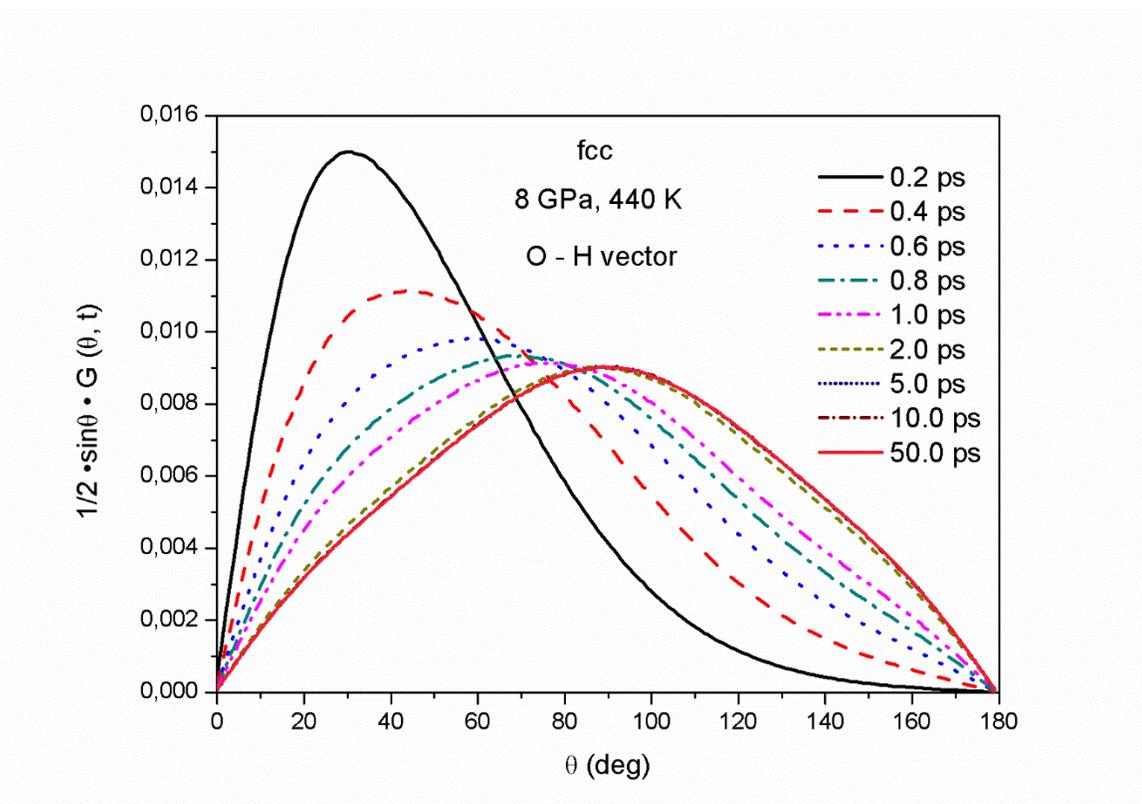

**Figure 11**

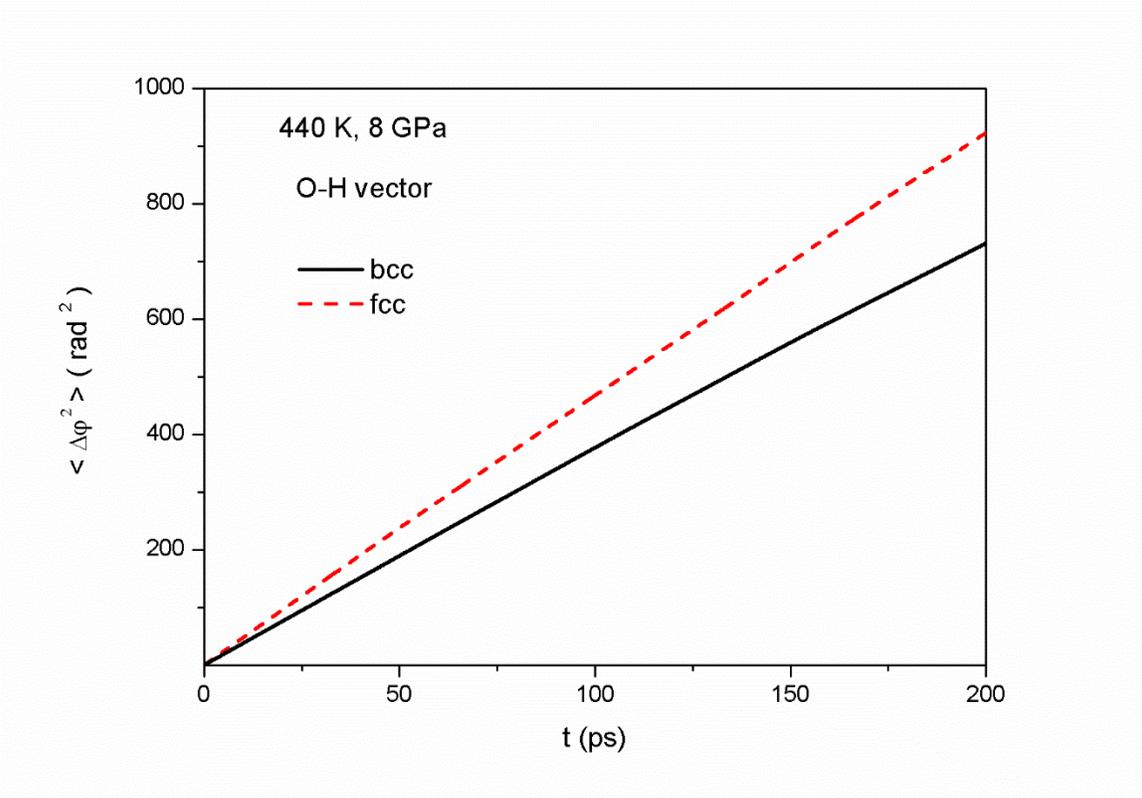

**Figure 12**

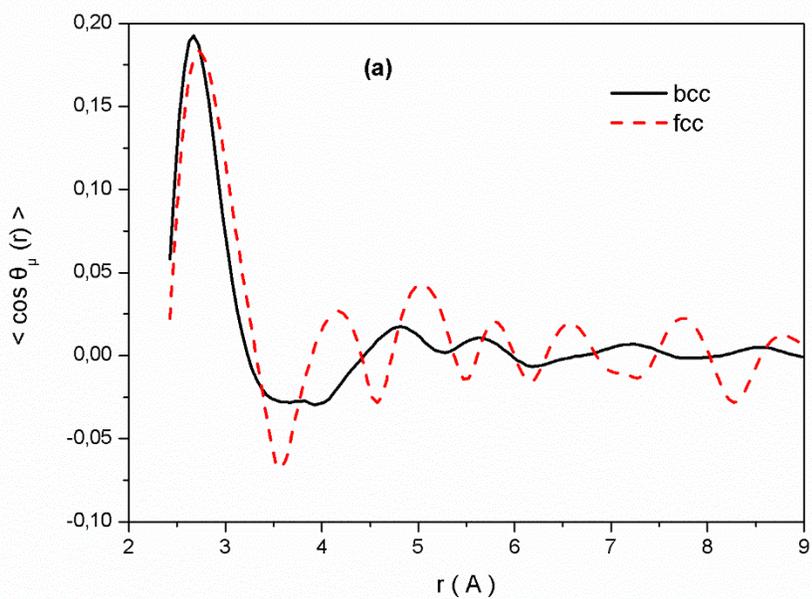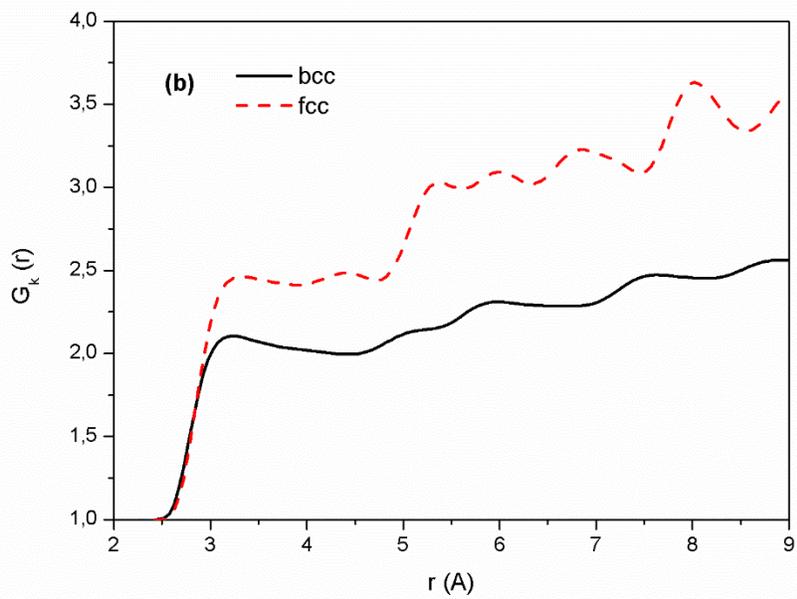

**Figure 13**

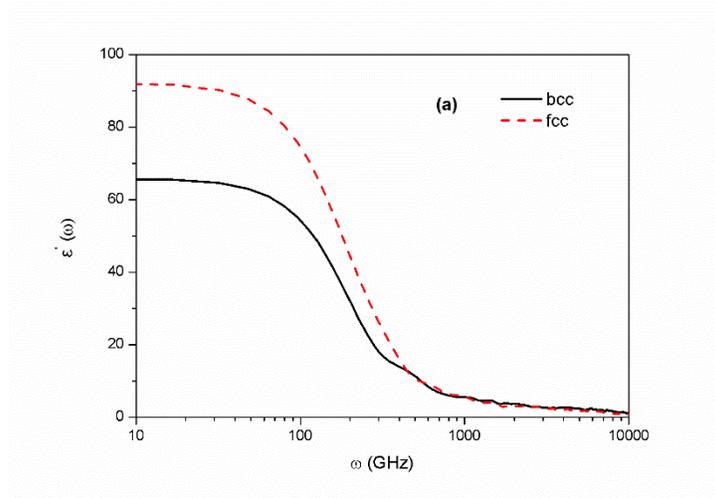

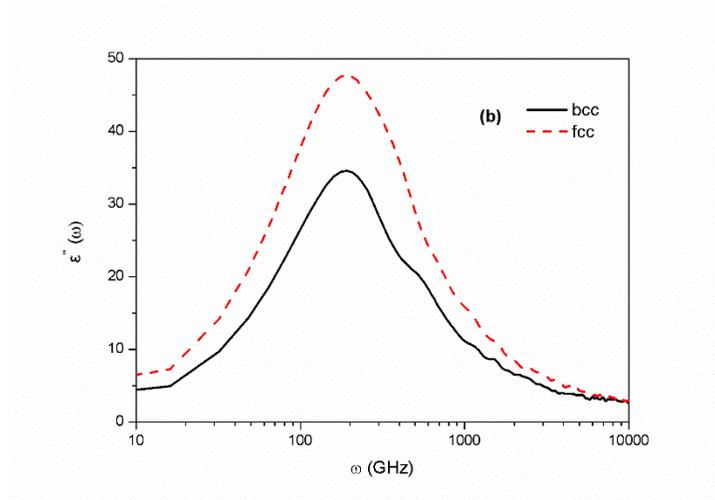

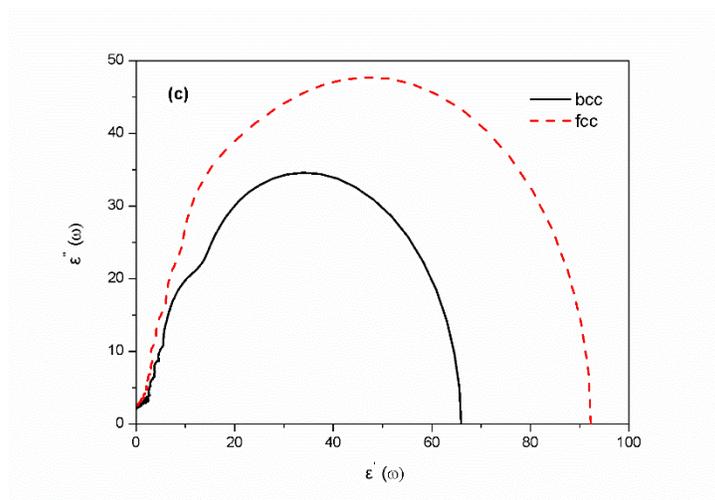

**Figure 14**

## Figure Captions

**Figure 1:** The calculated O-O, O-H and H-H rdfs, corresponding to the bcc and fcc plastic crystal phases at 440 K and 8 GPa.

**Figure 2:** Contributions of the six nearest neighbors to the overall O-O rdf in both the fcc and bcc plastic crystal phases.

**Figure 3:** The calculated continuous HB tcfs $C_{HB}^{C}(t)$ for the bcc and fcc plastic crystal phases at 440 K and 8 GPa.

**Figure 4:** The calculated distribution of the structural descriptor $\zeta$ for both the fcc and bcc plastic crystal phases.

**Figure 5:** The calculated distributions for the $q_4$ and $q_3$ orientational order parameters for both the fcc and bcc plastic crystal phases.

**Figure 6:** The calculated tcfs of the $q_4$ and $q_3$ orientational order parameters for both the fcc and bcc plastic crystal phases.

**Figure 7:** The calculated neutron and X-Ray structure factors for both the fcc and bcc plastic crystal phases.

**Figure 8:** The calculated oxygen and hydrogen atomic velocity tcfs for both the fcc and bcc plastic crystal phases.

**Figure 9:** The calculated spectral densities corresponding to the oxygen and hydrogen atomic velocity tcfs for both the fcc and bcc plastic crystal phases.

**Figure 10:** The calculated Legendre reorientational tcfs corresponding to the O-H vector of the water molecules for both the fcc and bcc plastic crystal phases.

**Figure 11:** The calculated $1/2 \cdot \sin\theta \cdot G(\theta, t)$ functions corresponding to the O-H vector of the water molecules for both the fcc and bcc plastic crystal phases.

**Figure 12:** The calculated angular mean square displacement of the O-H vector of the water molecules as a function of time for both the fcc and bcc plastic crystal phases.

**Figure 13:** (a) The calculated average value of the cosine of the angle formed by the dipole moment vectors of two water molecules as a function of their distance r. (b) The calculated radial dependence of the finite system Kirkwood factor, as a function of the distance around each water molecule.

**Figure 14:** (a) The calculated real part of the frequency dependent dielectric constant for both the fcc and bcc plastic crystal phases. (b) The calculated imaginary part of the frequency dependent dielectric constant for both the fcc and bcc plastic crystal phases. (c) The calculated Cole-Cole plots for both the fcc and bcc plastic crystal phases.